\documentclass[american,english]{IEEEtran}
\usepackage[T1]{fontenc}
\usepackage[latin9]{inputenc}
\usepackage[a4paper]{geometry}
\usepackage{color}
\usepackage{amsmath}
\usepackage{amsthm}
\usepackage{amssymb}
\usepackage{stackrel}
\usepackage{graphicx}
\usepackage{esint}

\makeatletter
\theoremstyle{plain}
\newtheorem{thm}{\protect\theoremname}
\theoremstyle{definition}
\newtheorem{defn}[thm]{\protect\definitionname}

\usepackage{colortbl}
\usepackage{cite}
\usepackage{nomencl}
\usepackage{flushend}
\usepackage{stfloats}
\usepackage[margin=8pt,font=footnotesize,labelfont=md,labelsep=colon]{caption}
\usepackage{amssymb}
\usepackage{amsmath}
\usepackage{mathtools}

\@ifundefined{showcaptionsetup}{}{%
 \PassOptionsToPackage{caption=false}{subfig}}
\usepackage{subfig}
\makeatother

\usepackage{babel}
\addto\captionsamerican{\renewcommand{\definitionname}{Definition}}
\addto\captionsamerican{\renewcommand{\theoremname}{Theorem}}
\addto\captionsenglish{\renewcommand{\definitionname}{Definition}}
\addto\captionsenglish{\renewcommand{\theoremname}{Theorem}}
\providecommand{\definitionname}{Definition}
\providecommand{\theoremname}{Theorem}

\begin{document}

\title{Full-Duplex Energy-Harvesting Enabled Relay Networks in Generalized
Fading Channels }

\author{\textcolor{black}{\normalsize{}Khaled Rabie, }\textit{\textcolor{black}{\normalsize{}Member,
IEEE}}\textcolor{black}{\normalsize{}, Bamidele Adebisi, }\textit{\textcolor{black}{\normalsize{}Senior
Member, IEEE}}\textcolor{black}{\normalsize{}, Galymzhan Nauryzbayev,
}\textit{\textcolor{black}{\normalsize{}Member, IEEE}}\textcolor{black}{\normalsize{},
}\\
\textcolor{black}{\normalsize{}Osamah S. Badarneh, }\textit{\textcolor{black}{\normalsize{} Member,
IEEE}}\textcolor{black}{\normalsize{}, Xingwang Li, }\textit{\textcolor{black}{\normalsize{}Member,
IEEE}}\textcolor{black}{\normalsize{}, Mohamed-Slim Alouini, }\textit{\textcolor{black}{\normalsize{}Fellow,
IEEE}}\textcolor{black}{\small{}}\\
\foreignlanguage{american}{\textcolor{black}{\small{}}}\thanks{\textcolor{black}{K. Rabie and B. Adebisi are with }\textcolor{black}{\small{}school
of Engineering, Manchester Metropolitan University, UK}\textcolor{black}{{}
(e-mails: }\textcolor{black}{\footnotesize{}k.rabie@mmu.ac.uk; b.adebisi@mmu.ac.uk}\textcolor{black}{).
}\protect \\
\textcolor{black}{G. }\textcolor{black}{\small{}Nauryzbayev}\textcolor{black}{{}
is with }\textcolor{black}{\small{}the College of Science and Engineering,
Hamad Bin Khalifa University, Qatar Foundation, Doha, Qatar}\textcolor{black}{{}
(e-mail: }\textcolor{black}{\footnotesize{}gnauryzbayev@hbku.edu.qa}\textcolor{black}{).}\protect \\
\textcolor{black}{\small{}O. Badarneh}\textcolor{black}{{} is with }\textcolor{black}{\small{}xxxxxxxxxxxxxxxxxxxxxxxxxxxxxxxxxx}\textcolor{black}{.}\protect \\
\textcolor{black}{X. Li is with }\textcolor{black}{\small{}school
of Physical and Electronics Engineering, Henan Polytechnic University
(HPU), China}\textcolor{black}{{} (e-mail: }\textcolor{black}{\footnotesize{}lixingwang@hpu.edu.cn}\textcolor{black}{).}\textcolor{black}{\small{}}\protect \\
\textcolor{black}{\small{}Mohamed-Slim Alouini}\textcolor{black}{{}
is with }\textcolor{black}{\small{}King Abdullah University of Science
and Technology (KAUST), Thuwal, Mekkah Province, Saudi Arabia}\textcolor{black}{{}
(e-mail: }\textcolor{black}{\footnotesize{}slim.alouini@kaust.edu.sa}\textcolor{black}{).}}}

\maketitle
\selectlanguage{american}%
\textcolor{black}{\thispagestyle{empty}
\pagestyle{empty}}
\selectlanguage{english}%
\begin{abstract}
\textcolor{black}{This paper analyzes the performance of a full-duplex
decode-and-forward relaying network over the generalized $\kappa$-$\mu$
fading channel. The relay is energy-constrained and relies entirely
on harvesting the power signal transmitted by the source based on
the time-switching relaying protocol. A unified analytical expression
for the ergodic outage probability is derived for the system under
consideration. This is then used to derive closed-form analytical
expressions for three special cases of the $\kappa$-$\mu$ fading
model, namely, Nakagami-$m$, Rice and Rayleigh. Monte Carlo simulations
are provided throughout to verify the correctness of our analysis. }\end{abstract}

\begin{IEEEkeywords}
Decode-and-forward (DF) relaying, energy-harvesting, full-duplex (FD),
generalized $\kappa$$-$$\mu$ fading.
\end{IEEEkeywords}

\section{\label{sec:Introduction}Introduction}

\selectlanguage{american}%
\IEEEPARstart{S}{imultaneous}\foreignlanguage{english}{ wireless
information and power transfer (SWIPT) in full-\textcolor{black}{duplex
(FD) relaying networks has recently attracted a great deal of research
attention. For instance, the authors in \cite{swiptFD14} considered
a dual-hop FD SWIPT network with both amplify-and-forward (AF) and
decode-and-forward (DF) relaying protocols equipped with a single
antenna. Several analytical expressions of the achievable throughput
were derived. Instead of the single-antenna relay, the authors in
\cite{swiptFD15MIMO} extended the work in \cite{swiptFD14} to include
multiple-input multiple-output (MIMO) FD relaying. Unlike \cite{swiptFD14}
and \cite{swiptFD15MIMO}, which focused on Rayleigh fading channels,
the work in \cite{RabieieeeTGCN1} analyzed the performance of a FD
SWIPT system in indoor environments characterized by log-normal fading.
The study in \cite{GalymFD} considered the outage probability of
FD SWIPT networks over $\alpha-\mu$ fading channels. Furthermore,
the authors in \cite{swiptFD17} studied physical layer security in
FD two-way relaying for SWIPT. More specifically, AF relaying with
multiple antennas and zero-forcing was deployed in this work. Other
studies exploiting jamming signals for EH have recently appeared in
\cite{EH_AN1,EH_AN2}. }}

\selectlanguage{english}%
\textcolor{black}{None of the aforementioned works considered FD energy-harvesting
(EH)-enabled relay networks over generalized fading channels. In contrast,
and motivated by this lack of analytical analysis, we present in this
letter a thorough performance evaluation of FD EH-enabled relay networks
over the generalized $\kappa$-$\mu$ fading channel. Specifically,
DF and time-switching relaying (TSR) protocols are deployed at the
relay. The motivations of our work come from the following two factors.
Firstly, the $\kappa$-$\mu$ model is a small-scale fading model
and is able to characterize the scattering cluster in homogeneous
communication environments including Rice $\left(\kappa=k,\,\mu=1\right)$,
Nagakami-$m$ $\left(\kappa\rightarrow0,\,\mu=m\right)$ and Rayleigh
$\left(\kappa\rightarrow0,\,\mu=1\right)$ \cite{KappaMu13,KappaMu15,KappaMu16}.
Secondly, FD communication allows devices to operate on the same frequency,
which potentially doubles the spectral efficiency, and, because of
this, FD has become a viable option for next generation wireless communication
networks. Thus, the network studied herein is meaningful and valuable
for consideration. }

\textcolor{black}{The main contribution of this letter resides in
deriving a novel unified analytical expression for the ergodic outage
probability of the proposed system over the generalized $\kappa$-$\mu$
fading channel. In addition, closed-form expressions for the aforementioned
special cases of the $\kappa$-$\mu$ fading scenario are presented.
The derived expressions were used to investigate the impact of several
system and fading parameters on the performance. Results show that
the performance can be enhanced considerably as the fading parameters
$\kappa$ and $\mu$ are increased. It is also shown that as the }loop-back
interference, associated with FD relaying, increases the outage probability
performance deteriorates drastically.

\section{System Model\label{sec:System-Mode}}

The considered system model consists of a source (S), a relay (R)
and a destination (D). The end nodes are equipped wi\textcolor{black}{th
a single antenna whereas R, based on DF, has two antennas operating
in FD mode. It is assumed that there is no direct link between the
end nodes due to severe shadowing and path-loss effect \cite{swiptFD14,swiptFD15MIMO,RabieieeeTGCN1,swiptFD17};
hence, all communication is accomplished over two phases. The S-to-R,
R-to-D and loop-back interference channel coefficients, denoted as
$h_{1}$, $h_{2}$ and $h_{3}$, are assumed to be independent but
not necessarily identical following the $\kappa$-$\mu$ distribution
with a probability density function (PDF) 
\begin{eqnarray}
f_{Z_{i}}\left(z\right)=\Upsilon_{i}\,z^{\frac{\mu_{i}-1}{2}}\,\textrm{exp}\left(-\frac{\phi_{i}\,z}{\Omega_{i}}\right)I_{\mu_{i}-1}\left(2\mu_{i}\sqrt{\frac{\phi_{i}\,z}{\Omega_{i}}}\right),\label{eq:pdfKappa}
\end{eqnarray}
}

\noindent \textcolor{black}{where $i\in\left\{ 1,2,3\right\} $, $Z_{i}=h_{i}^{2},$
$\Omega_{i}=\mathbb{E}\left[Z_{i}\right]$, $\phi_{i}=\mu_{i}\left(1+\kappa_{i}\right),$
$\Upsilon_{i}=\frac{\mu_{i}\left(1+\kappa_{i}\right)^{\frac{\mu_{i}+1}{2}}}{\textrm{exp}\left(\mu_{i}\kappa_{i}\right)\kappa_{i}^{\frac{\mu_{i}-1}{2}}\Omega_{i}^{\frac{\mu_{i}+1}{2}}}$,
$I_{v}\left[\cdot\right]$ is the modified Bessel function of the
first kind with arbitrary order $v$ \cite[Eq. 9.6.20]{BookAbramow72},
$\mu_{i}$ represents the number of the multipath clusters and $\kappa_{i}>0$
denotes the ratio between the total powers of the domain components
and the scattered waves. The path-loss exponents for the S-to-R and
R-to-D links are denoted by $\xi_{1}$ and $\xi_{2}$, respectively.
It is assumed that perfect channel state information (CSI) is available
at all receiving nodes. R has no power supply and operates by harvesting
the RF signal coming from S. The energy used for information processing
at R is negligible and hence all the harvested energy will be utilized
to forward the source information. }

\textcolor{black}{As mentioned earlier, the TSR protocol is used for
EH at R, in which the time frame }$T$ is divided into two consecutive
time slots: $\alpha T$ and $\left(1-\alpha\right)T$, which are used
for EH and  S-to-R / R-to-D information transmissions, respectively;
where $0\leq\alpha\leq1$ is the EH time factor. 

For the sake of brevity, we omit the mathematical modeling of the
received signals at R and D and present only the corresponding signal-to-noise
ratios (SNRs). Readers may refer t\textcolor{black}{o \cite{swiptFD14,RabieieeeTGCN1}
for m}ore details. The SNRs at R and D nodes are expressed respectively
as 

\begin{equation}
\gamma_{r}=\frac{P_{s}h_{1}^{2}}{P_{r}d_{1}^{\xi_{1}}h_{3}^{2}}=\frac{1}{\zeta h_{3}^{2}},\,\textrm{and}\label{eq:snrR}
\end{equation}

\begin{equation}
\gamma_{d}=\frac{P_{r}h_{2}^{2}}{d_{2}^{\xi_{2}}\sigma_{d}^{2}}=\frac{\zeta P_{s}h_{1}^{2}h_{2}^{2}}{d_{1}^{\xi_{1}}d_{2}^{\xi_{2}}\sigma_{d}^{2}},\label{eq:snrD}
\end{equation}

\noindent where $P_{s}$ is the source transmit power, $P_{r}=\eta\alpha P_{s}h_{1}^{2}/\left(\left(1-\alpha\right)d_{1}^{\xi_{1}}\right)$
is the relay transmit power, $\eta$ is efficiency of the energy harvester,
$\sigma_{d}^{2}$ is the noise variance at D, $\zeta=\frac{\eta\alpha}{1-\alpha}$,
$d_{1}$ and $d_{2}$ are the S-to-R and R-to-D distances, respectively. 

The instantaneous capacity of the first and second links can be given
by 

\begin{equation}
C_{i}=\left(1-\alpha\right)\textrm{log}{}_{2}\left(1+\gamma_{i}\right),\quad i\in\left\{ r,\,d\right\} .\label{eq:C}
\end{equation}

With this in mind, the ergodic outage probability, which is defined
as the probability that the instantaneous capacity falling below a
certain threshold $\left(C_{th}\right)$, can be calculated as 

\begin{equation}
P_{out}\left(C_{th}\right)=\Pr\left(\textrm{min}\left\{ C_{r},\,C_{d}\right\} <C_{th}\right).\label{eq:Pout-1}
\end{equation}

\section{\label{sec:Performance analysis}Performance Analysis }

\vspace{2mm}

In this section, we derive analytical expressions of the ergodic outage
capacity in generalized $\kappa$-$\mu$ fading and its special cases.
To begin with, we substitute (\ref{eq:snrR}) and (\ref{eq:snrD})
into (\ref{eq:C}) and then into (\ref{eq:Pout-1}), with some mathematical
manipulations, to obtain 

\begin{equation}
P_{out}\left(C_{th}\right)=\Pr\left(\textrm{min}\left\{ \frac{1}{\zeta Z},\,\frac{\zeta P_{s}W}{d_{1}^{\xi_{1}}d_{2}^{\xi_{2}}\sigma_{d}^{2}}\right\} <\upsilon\right),\label{eq:Pout-1-1}
\end{equation}

\noindent where $X=h_{1}^{2},$ $Y=h_{2}^{2}$, $Z=h_{3}^{2}$, $W=XY$
and $\upsilon=2^{\frac{Cth}{1-\alpha}}-1$.

Because t\textcolor{black}{he random variables (RVs) $W$ and $Z$
are independent, we can calculate the probability in (\ref{eq:Pout-1-1})
as}

\textcolor{black}{
\begin{equation}
P_{out}\left(C_{th}\right)=1-\bar{F}_{Z}\left(\frac{1}{\zeta\upsilon}\right)\bar{F}_{W}\left(\frac{d_{1}^{\xi_{1}}d_{2}^{\xi_{2}}\sigma_{d}^{2}}{\zeta P_{s}}\upsilon\right).\label{eq:Poutfinal}
\end{equation}
}

\noindent \textcolor{black}{where $\bar{F}_{Z}\left(\cdot\right)$
is the complementary cumulative distribution function (CCDF) of $Z$,
which can be obtained by integrating (\ref{eq:pdfKappa}) with appropriate
notation changes. In order to arrive at a tractable expression, we
use the series representation of $I_{\mu_{3}-1}\left(\cdot\right)$
\cite[eq.  8.445]{book2}; that is }

\textcolor{black}{
\begin{equation}
I_{\mu_{3}-1}\left(2\Lambda\right)=\stackrel[q=0]{\infty}{\sum}\frac{1}{\Gamma\left(\mu_{3}+q\right)q!}\,\Lambda^{\mu_{3}-1+2q},\label{eq:Iv_series}
\end{equation}
}

\noindent \textcolor{black}{where $\Lambda=\mu_{3}\sqrt{\kappa_{3}\left(1+\kappa_{3}\right)z}$
and $\Gamma\left(\cdot\right)$ is the Gamma function \cite[Eq. 8.310.1]{BookGrad2000}. }

\textcolor{black}{Now, replacing (\ref{eq:Iv_series}) in (\ref{eq:pdfKappa})
and then integrating, we can express the cumulative distribution function
(CDF) $F_{Z}\left(\upsilon\right)$ a}s
\begin{eqnarray}
F_{Z}\left(\upsilon\right)=\Upsilon_{3}\stackrel[q=0]{\infty}{\sum}\frac{\left(\mu_{3}\sqrt{\kappa_{3}\left(1+\kappa_{3}\right)}\right)^{\mu_{3}-1+2q}}{\Gamma\left(\mu_{3}+q\right)q!}\,I_{0},\label{eq:FZ}
\end{eqnarray}

\noindent where $b=\frac{1-\alpha}{\eta\alpha}$, $\Upsilon_{3}=\frac{\mu_{3}\left(1+\kappa_{3}\right)^{\frac{\mu_{3}+1}{2}}\kappa_{3}^{-\frac{\mu_{3}-1}{2}}}{\textrm{exp}\left(\mu_{3}\kappa_{3}\right)}$
and 
\begin{eqnarray}
I_{0} & = & \stackrel[0]{\frac{b}{\upsilon}}{\int}z^{\mu_{3}+q-1}\textrm{ exp}\left(-\phi_{3}z\right)\textrm{d}z\nonumber \\
 & \stackrel{\left(a\right)}{=} & \phi_{3}^{-\mu_{3}-q}\gamma\left(\mu_{3}+q,\frac{\phi_{3}b}{\upsilon}\right),\label{eq:I0}
\end{eqnarray}
\textcolor{black}{where $\gamma\left(\cdot,\cdot\right)$ is the lower
incomplete Gamma function \cite[Eq. 8.350.1]{BookGrad2000}. Note
that $\left(a\right)$ is obtained with the help of \cite[Eq. 3.351.1]{book2}. }

Substituting (\ref{eq:I0}) into (\ref{eq:FZ}), along with some straightforward
manipulations, yields
\begin{eqnarray}
F_{Z}\left(\upsilon\right)=\frac{1}{\textrm{exp}\left(\kappa_{3}\mu_{3}\right)}\stackrel[q=0]{\infty}{\sum}\frac{\left(\mu_{3}\kappa_{3}\right)^{q}}{\Gamma\left(\mu_{3}+q\right)q!}\gamma\left(\mu_{3}+q,\frac{\phi_{3}b}{\upsilon}\right).\label{eq:FZfinal}
\end{eqnarray}

\begin{defn}
For any two independent RVs $U$ and $V$, the CDF of the product
of them is defined as $P\left(UV\leq x\right)=\int F_{V}\left(\frac{x}{u}\right)f_{U}\left(u\right)\textrm{d}u$.

Using this definition, we can determine the CDF of $W$ as
\begin{equation}
F_{W}\left(r\right)=\stackrel[0]{\infty}{\int}F_{X}\left(\frac{r}{u}\right)f_{Y}\left(u\right)\textrm{d}u,\label{eq:FX}
\end{equation}
where $f_{Y}\left(u\right)$ has the distribution as in (\ref{eq:pdfKappa})
and $F_{X}\left(\frac{r}{u}\right)$ can be obtained from (\ref{eq:FZfinal})
with the appropriate change of notations. With this in mind, (\ref{eq:FX})
can be expressed as{\small{}
\begin{eqnarray}
F_{W}\left(\upsilon\right)=\frac{\Upsilon_{2}}{\textrm{exp}\left(\kappa_{1}\mu_{1}\right)}\stackrel[n=0]{\infty}{\sum}\frac{\left(\kappa_{1}\mu_{1}\right)^{n}}{\Gamma\left(\mu_{1}+n\right)n!}\Biggl[\Gamma\left(\mu_{1}+n\right)I_{1}-I_{2}\Biggr],\label{eq:FW}
\end{eqnarray}
}where $\Upsilon_{2}=\frac{\mu_{2}\left(1+\kappa_{2}\right)^{\frac{\mu_{2}+1}{2}}\kappa_{2}^{-\frac{\mu_{2}-1}{2}}}{\textrm{exp}\left(\mu_{2}\kappa_{2}\right)}$,
\begin{equation}
I_{1}=\stackrel[0]{\infty}{\int}z^{\frac{\mu_{2}-1}{2}}\textrm{exp}\left(-\phi_{2}z\right)\,I_{\mu_{2}-1}\left(2\Delta_{2}\sqrt{z}\right)\textrm{d}z,
\end{equation}
\begin{equation}
I_{2}=\stackrel[0]{\infty}{\int}z^{\frac{\mu_{2}-1}{2}}\textrm{exp}\left(-\phi_{2}z\right)I_{\mu_{2}-1}\left(2\Delta_{2}\sqrt{z}\right)\Gamma\left(\mu_{1}+n,\frac{\phi_{1}\upsilon}{az}\right)\textrm{d}z,
\end{equation}

\end{defn}
\noindent \textcolor{black}{$\Delta_{2}=\mu_{2}\sqrt{\kappa_{2}\left(1+\kappa_{2}\right)}$,
$a=\frac{\eta\alpha P_{s}}{\left(1-\alpha\right)d_{1}^{m}d_{2}^{m}\sigma_{d}^{2}}$
and $\Gamma\left(\cdot,\cdot\right)$ indicates the upper incomplete
Gamma function \cite[Eq. 8.350.2]{BookGrad2000}.}

\textcolor{black}{To the best of the authors' knowledge, there is
no analytical solution for the integral $I_{1}$. Hence, to solve
this integral, we use the infinite series representation of $I_{\mu_{2}-1}\left(\cdot\right)$
\cite[Eq.  8.445]{book2}. Thus, $I_{1}$ can be rewritten as }

\textcolor{black}{
\begin{eqnarray}
I_{1} & = & \stackrel[m=0]{\infty}{\sum}\frac{\Delta_{2}^{\mu_{2}+2m-1}}{\Gamma\left(\mu_{2}+m\right)m!}\stackrel[0]{\infty}{\int}z^{\mu_{2}+m-1}\textrm{exp}\left(-\phi_{2}z\right)\textrm{d}z\nonumber \\
 & \stackrel{\left(a\right)}{=} & \stackrel[m=0]{\infty}{\sum}\frac{\left(\mu_{2}\sqrt{\kappa_{2}\left(\kappa_{2}+1\right)}\right)^{\mu_{2}+2m-1}}{\phi_{2}^{\mu_{2}+m}\,m!},\label{eq:I1}
\end{eqnarray}
}

\noindent \textcolor{black}{where $\left(a\right)$ is obtained with
the help of \cite[Eq. 3.351.3]{book2} along with some }basic algebraic
manipulations.

Similarly, to solve the integral $I_{2}$, we first replace $I_{\mu_{2}-1}\left(\cdot\right)$
and $\Gamma\left(\cdot,\cdot\right)$ with their series representations
using \cite[Eq.  8.445]{book2} and \cite[Eq.  8.352.2]{book2}, respectively,
as follows

\begin{equation}
I_{\mu_{2}-1}\left(2\Delta\right)=\stackrel[l=0]{\infty}{\sum}\frac{1}{\Gamma\left(\mu_{2}+l\right)l!}\,\left(\Delta_{2}\sqrt{z}\right)^{\mu_{2}-1+2l},\label{eq:Iv_series-1}
\end{equation}

\begin{eqnarray}
\Gamma\left(\mu_{1}+n,\frac{\phi_{1}\upsilon}{a\,z}\right)=\psi!\textrm{ exp}\left(-\frac{\phi_{1}\upsilon}{a\,z}\right)\stackrel[k=0]{\psi}{\sum}\frac{1}{k!}\left(\frac{\phi_{1}\upsilon}{a\,z}\right)^{k},\label{eq:Iv_series-1-1}
\end{eqnarray}

\noindent where $\psi=\mu_{1}+n-1$. 

Using (\ref{eq:Iv_series-1}) and (\ref{eq:Iv_series-1-1}), we can
rewrite $I_{2}$ as 
\begin{eqnarray}
\nonumber \\
I_{2}=\psi\stackrel[l=0]{\infty}{\sum}\,\stackrel[k=0]{\psi}{\sum}\left(\frac{\phi_{1}\upsilon}{a}\right)^{k}\frac{\Delta_{2}^{\mu_{2}+2l-1}}{\Gamma\left(\mu_{2}+l\right)\,l!\,k!}\,I_{3},\label{eq:I2}
\end{eqnarray}

\noindent where 

\begin{eqnarray}
I_{3} & = & \stackrel[0]{\infty}{\int}z^{\mu_{2}+l-k-1}\textrm{exp}\left(-\phi_{2}\,z-\frac{\phi_{1}\upsilon}{a\,z}\right)\textrm{d}z\nonumber \\
 & \stackrel{\left(a\right)}{=} & 2\left(\frac{\phi_{2}\,a}{\phi_{1}\upsilon}\right)^{\frac{1}{2}\left(k-l-\mu_{2}\right)}K_{k-l-\mu_{2}}\left(\sqrt{\frac{\upsilon}{a}\phi_{1}\phi_{2}}\right),\quad\label{eq:I3}
\end{eqnarray}

\noindent 
\begin{figure*}[t]
\begin{eqnarray}
F_{W}\left(\upsilon\right)=\frac{\mu_{2}\left(1+\kappa_{2}\right)^{\frac{\mu_{2+1}}{2}}}{\kappa_{2}^{\frac{\mu_{2}-1}{2}}\underset{i\in\left\{ 1,2\right\} }{\prod}\textrm{exp}\left(\kappa_{i}\mu_{i}\right)}\Biggl\{\stackrel[n=0]{\infty}{\sum}\,\stackrel[m=0]{\infty}{\sum}\frac{\left(\mu_{2}\sqrt{\kappa_{2}\left(1+\kappa_{2}\right)}\right)^{\mu_{2}+2m-1}\left(\kappa_{1}\mu_{1}\right)^{n}}{\left(\left(1+\kappa_{2}\right)\mu_{2}\right)^{\mu_{2}+m}m!\,n!}-2\stackrel[n=0]{\infty}{\sum}\,\stackrel[l=0]{\infty}{\sum}\stackrel[k=0]{\psi}{\sum}\nonumber \\
\frac{\left(\kappa_{1}\mu_{1}\right)^{n}\left(\mu_{1}+n-1\right)!}{n!\,k!\,l!}\frac{\left(\mu_{2}\sqrt{\kappa_{2}\left(1+\kappa_{2}\right)}\right)^{\mu_{2}+2l-1}}{\Gamma\left(\mu_{1}+n\right)\Gamma\left(\mu_{2}+l\right)}\left(\frac{\phi_{1}\upsilon}{a}\right)^{k}\left(\frac{a\phi_{2}}{\upsilon\phi_{1}}\right)^{-\frac{1}{2}\left(\mu_{2}+l-k\right)}\,K_{k-l-\mu_{2}}\left[\,2\sqrt{\frac{\upsilon}{a}}\sqrt{\phi_{1}\phi_{2}}\right]\Biggr\}\label{eq:FWfinal}
\end{eqnarray}

\selectlanguage{american}%
\centering{}\rule[0.5ex]{2.03\columnwidth}{0.8pt}\selectlanguage{english}%
\end{figure*}
\begin{figure*}[t]
\begin{eqnarray}
P_{out}\left(\upsilon\right)=1-\frac{1}{\textrm{exp}\left(\kappa_{3}\mu_{3}\right)}\stackrel[q=0]{\infty}{\sum}\left\{ \frac{\left(\mu_{3}\kappa_{3}\right)^{q}}{\Gamma\left(\mu_{3}+q\right)q!}\,\gamma\left(\mu_{3}+q,\frac{\phi_{3}b}{\upsilon}\right)\right\} \Biggl(1-\frac{\mu_{2}\left(1+\kappa_{2}\right)^{\frac{\mu_{2+1}}{2}}}{\kappa_{2}^{\frac{\mu_{2}-1}{2}}\underset{i\in\left\{ 1,2\right\} }{\prod}\textrm{exp}\left(\kappa_{i}\mu_{i}\right)}\Biggl\{\stackrel[n=0]{\infty}{\sum}\stackrel[m=0]{\infty}{\sum}\frac{\Delta_{2}^{\mu_{2}+2m-1}}{\phi_{2}^{\mu_{2}+m}}\nonumber \\
\times\frac{\left(\kappa_{1}\mu_{1}\right)^{n}}{m!\,n!}-2\stackrel[n=0]{\infty}{\sum}\stackrel[l=0]{\infty}{\sum}\stackrel[k=0]{\psi}{\sum}\frac{\left(\kappa_{1}\mu_{1}\right)^{n}\Delta_{2}^{\mu_{2}+2l-1}\psi!}{\Gamma\left(\mu_{1}+n\right)\Gamma\left(\mu_{2}+l\right)n!\,k!\,l!}\left(\frac{\phi_{1}\upsilon}{a}\right)^{k}\left(\frac{\phi_{2}a}{\phi_{1}\upsilon}\right)^{-\frac{1}{2}\left(\mu_{2}+l-k\right)}K_{k-l-\mu_{2}}\left[\,2\sqrt{\frac{\upsilon}{a}\phi_{1}\phi_{2}}\right]\Biggr\}\Biggr)\label{eq:Poutfinal-1}
\end{eqnarray}

\selectlanguage{american}%
\centering{}\rule[0.5ex]{2.03\columnwidth}{0.8pt}\selectlanguage{english}%
\end{figure*}
\begin{figure*}[t]
\begin{eqnarray}
P_{out}^{\{\textrm{Ric}\}}\left(C_{th}\right)=1-\frac{1}{\textrm{exp}\left(\kappa_{3}\right)}\,\stackrel[q=0]{\infty}{\sum}\left\{ \frac{\kappa_{3}^{q}}{\Gamma\left(q+1\right)q!}\,\gamma\left(q+1,\,\left(1+\kappa_{3}\right)\frac{b}{\upsilon}\right)\right\} \Biggl(1-\frac{1}{\textrm{exp}\left(\kappa_{1}+\kappa_{2}\right)}\Biggl\{\stackrel[n=0]{\infty}{\sum}\,\stackrel[m=0]{\infty}{\sum}\frac{\kappa_{1}^{n}\kappa_{2}^{m}}{n!\,m!}\nonumber \\
-2\stackrel[n=0]{\infty}{\sum}\,\stackrel[l=0]{\infty}{\sum}\,\stackrel[k=0]{n}{\sum}\frac{\kappa_{1}^{n}\kappa_{2}^{l}}{n!\,l!\,k!\,l!}\left(\left(\kappa_{1}+1\right)\left(\kappa_{1}+1\right)\frac{\upsilon}{a}\right)^{\frac{1}{2}\left(l+k+1\right)}\,K_{k-l-1}\left[2\sqrt{\left(1+\kappa_{1}\right)\left(1+\kappa_{2}\right)\frac{C_{th}}{a}}\right]\Biggr\}\Biggr)\label{eq:PoutR}
\end{eqnarray}

\selectlanguage{american}%
\centering{}\rule[0.5ex]{2.03\columnwidth}{0.8pt}\selectlanguage{english}%
\end{figure*}

\noindent where $\left(a\right)$ is accomplished by means of \cite[Eq. 3.471.12]{book2},
along with some mathematical manipulations. 

Substituting (\ref{eq:I3}) into (\ref{eq:I2}) and\textcolor{black}{{}
then (\ref{eq:I1}) and (\ref{eq:I2}) into (\ref{eq:FW}), we obtain
an expression for $F_{W}\left(\cdot\right)$ given in (\ref{eq:FWfinal}),
shown at the top of this page, where $K_{v}\left[\cdot\right]$ is
the modified Bessel function of the second kind with arbitrary order
$v$ \cite[Eq. 9.6.22]{BookAbramow72}. Finally, using (\ref{eq:Poutfinal}),
(\ref{eq:FZfinal}) and (\ref{eq:FWfinal}), along with} basic mathematical
manipulations, we obtain an accurate and unified expression for the
ergodic outage probability of the dual-hop FD-DF relaying system over
the g\textcolor{black}{eneralized $\kappa$-$\mu$ fading channel.
This is given by (\ref{eq:Poutfinal-1}), shown at the top of this
page. }

\textcolor{black}{Now, substituting $\mu_{1}=\mu_{2}=\mu_{3}=1$ in
(\ref{eq:Poutfinal-1}), we get an analytical} expression of the outage
probability for the Rice fading sc\textcolor{black}{enario, given
in (\ref{eq:PoutR}), shown at the top of the page. To obtain a mathematical
expression for the Nakagami-$m$ fading case, we start from (\ref{eq:Poutfinal-1})
and substitute $\kappa_{1}=\kappa_{2}=\kappa_{3}\rightarrow0$, $\mu_{1}=m_{1}$,
$\mu_{2}=m_{2}$ and $\mu_{3}=m_{3}$. Note that due to the fact that
$\kappa_{1}=\kappa_{2}=\kappa_{3}\rightarrow0$, only the first terms
of all infinite series will have non-zero values, expect the last
summation. Thus, the resultant closed-form expression of the ergodic
outage probability in Nakagami-$m$ fading can be given as }

\textcolor{black}{
\begin{eqnarray}
\negthickspace\negthickspace\negthickspace\negthickspace\negthickspace\negthickspace\negthickspace\negthickspace\negthickspace\negthickspace\negthickspace\negthickspace P_{out}^{\{\textrm{Nak}\}}=1-\frac{2}{\Gamma\left(m_{2}\right)\Gamma\left(m_{3}\right)}\,\gamma\left(m_{3},\,m_{3}\frac{b}{\upsilon}\right)\qquad\hfill\quad\nonumber \\
\hfill\hfill\times\stackrel[k=0]{m_{1}-1}{\sum}\frac{1}{k!}\left(\frac{m_{1}m_{2}\upsilon}{a}\right)^{\frac{m_{2}+k}{2}}K_{m_{2}-k}\left[\,2\sqrt{m_{1}m_{2}\frac{\upsilon}{a}}\right]\label{eq:PoutN}
\end{eqnarray}
}

\textcolor{black}{When $m_{1}=m_{2}=m_{3}=1$, the expression in (\ref{eq:PoutN})
reduces to the Rayleigh fading scenario as }

\textcolor{black}{
\begin{eqnarray}
P_{out}^{\{\textrm{Ray}\}}=1-2\sqrt{\frac{\upsilon}{a}}\left(1-\textrm{exp}\left(-\frac{b}{\upsilon}\right)\right)K_{1}\left[\,2\sqrt{\frac{\upsilon}{a}}\right],\label{eq:PoutRay}
\end{eqnarray}
}

\textcolor{black}{We know that $K_{1}\left(z\right)$ can be approximated
by $1/z$ when $z\ll1.$ Based on this, at high SNR, (\ref{eq:PoutRay})
can be simplified to $P_{out}^{\{\textrm{Ray}\}}\approx\textrm{exp}\left(-\frac{1-\alpha}{\eta\alpha\upsilon}\right),$
which indicates that the system performance improves as we increase
$\eta$ and/or decrease $\upsilon$. }

\section{\label{sec:Numerical-Results}Numerical Results and Discussions }

\begin{figure}
\begin{centering}
\includegraphics[scale=0.55]{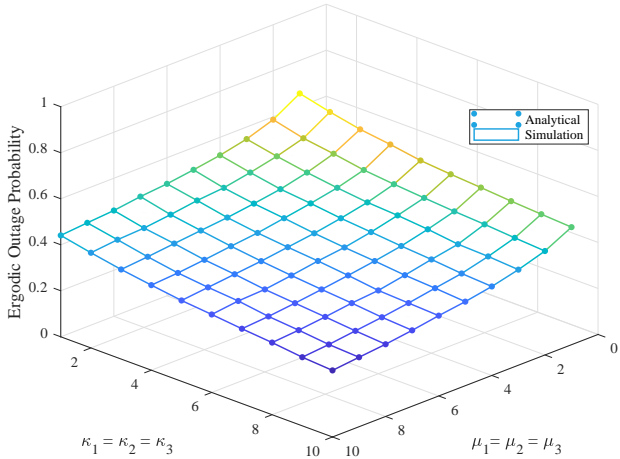}
\par\end{centering}

\caption{\label{fig:Fig1}Ergodic outage probability versus the fading parameters
$\kappa_{i}$ and $\mu_{i}$, $i\in\left\{ 1,2,3\right\} $, when
$\alpha=0.06$ and $P_{s}=0.5\textrm{W}$. }
\end{figure}

All our evaluations in this section, \textcolor{black}{unless we specify
otherwise, are based on: $\xi_{1}=\xi_{2}=\xi_{3}=2.7$, $d_{1}=d_{2}=4$m,
$\sigma_{r}=\sigma_{d}=0.01$W, $\eta=1$ and $C_{th}=0.2$bits/s/Hz
\cite{nasir}. To begin with, we show in Fig. \ref{fig:Fig1} a 3D
plot for the analytical and simulated outage probability as a function
of the fading parameters $\kappa$ and $\mu$. Note that the analytical
results are obtained using (\ref{eq:Poutfinal-1}) while considering
the first 20 terms of all series. It is clear that the performance
improves as $\kappa$ and/or $\mu$ is increase}d. This is because
increasing $\kappa$ indicates an increase in the ratio between the
total power of the dominant components and the total power of the
scattered waves, and increasing $\mu$ implies increasing the number
of multipath clusters. 

\begin{figure}
\begin{centering}
\subfloat[Nakagami-$m$ $\left(\kappa_{1}=\kappa_{2}\rightarrow0\right)$]{\begin{centering}
\includegraphics[scale=0.55]{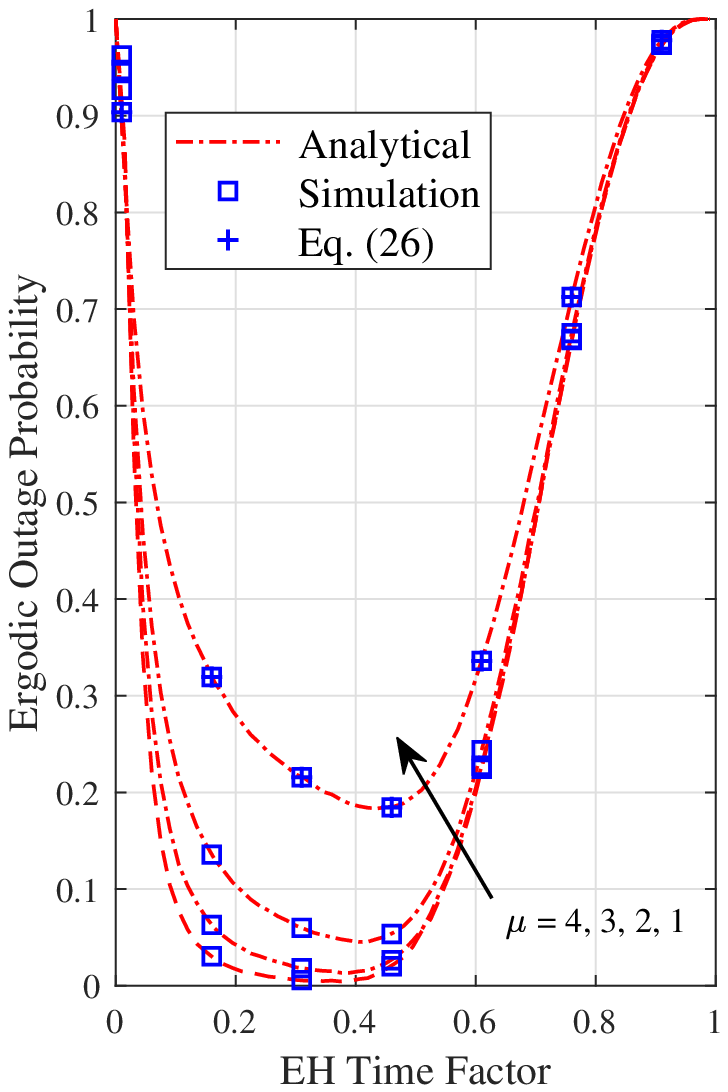}
\par\end{centering}

}\subfloat[Rice $\left(\mu_{1}=\mu_{2}=1\right)$]{\begin{centering}
\includegraphics[scale=0.55]{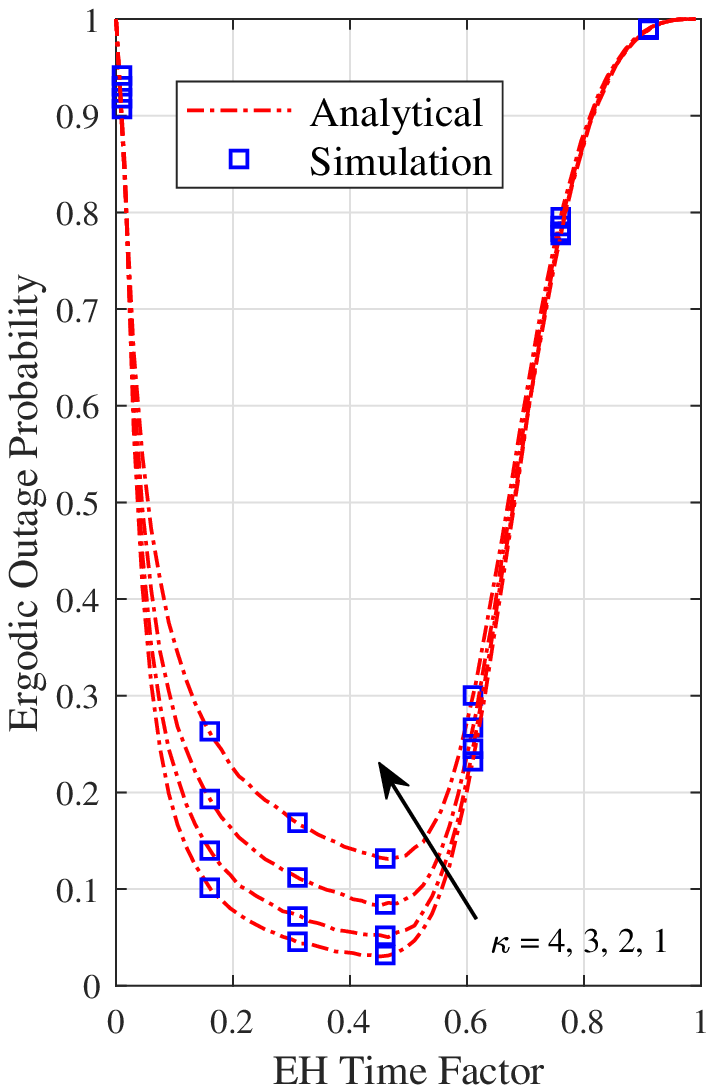}
\par\end{centering}

}
\par\end{centering}

\caption{\label{fig:Fig2}Ergodic outage probability versus $\alpha$ for the
FD-DF relay system for different $\kappa$ and $\mu$ values. }
\end{figure}

Now, to illustrate the influence of the EH time factor we present
in Fig. \ref{fig:Fig2} the ergodic outage probability with respect
to $\alpha$ for the two special cases of the $\kappa$-$\mu$ fading
model: Nakagami-$m$ fading in Fig. \ref{fig:Fig2}(a) for vario\textcolor{black}{us
values of $\mu$ and Rice fading in Fig. \ref{fig:Fig2}(b) with different
values of $\kappa$. Note that the numerical results in Figs. \ref{fig:Fig2}(a)
and \ref{fig:Fig2}(b) are obtained from (\ref{eq:PoutN}) and (\ref{eq:PoutR}),
respectively. It is clear that, for all fading scenarios, when $\alpha$
is either too high or too small, the performance degrades significantly;
hence, this parameter must be selected carefully to minimize the outage
probability. It is worthwhile pointing out that the results represented
by the symbol (+) in Fig. \ref{fig:Fig2}(a) are for Rayleigh fading
and are obtained from (\ref{eq:PoutRay}).}

\begin{figure}
\centering{}%
\begin{minipage}[t]{0.52\columnwidth}%
\begin{flushleft}
\includegraphics[scale=0.55]{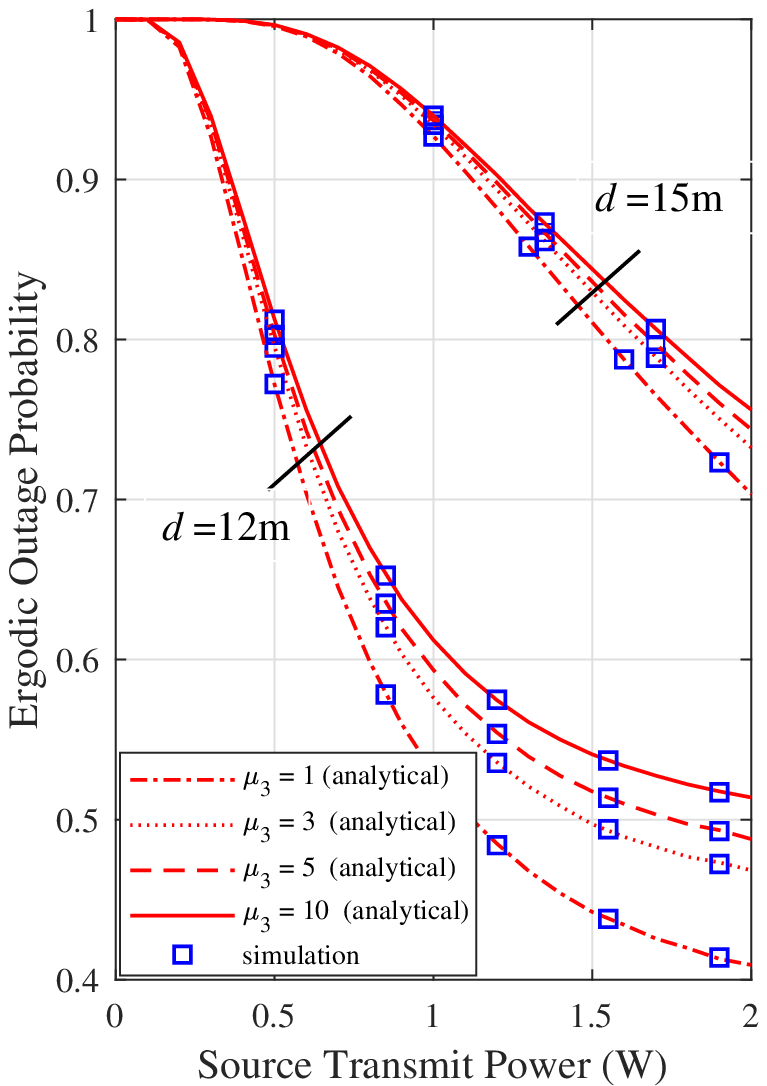}
\par\end{flushleft}

\caption{\label{fig:Fig3}Ergodic outage probability with respect to $P_{s}$
for different loop-back interference scenarios.}
\end{minipage}%
\begin{minipage}[t]{0.52\columnwidth}%
\begin{flushleft}
\includegraphics[scale=0.55]{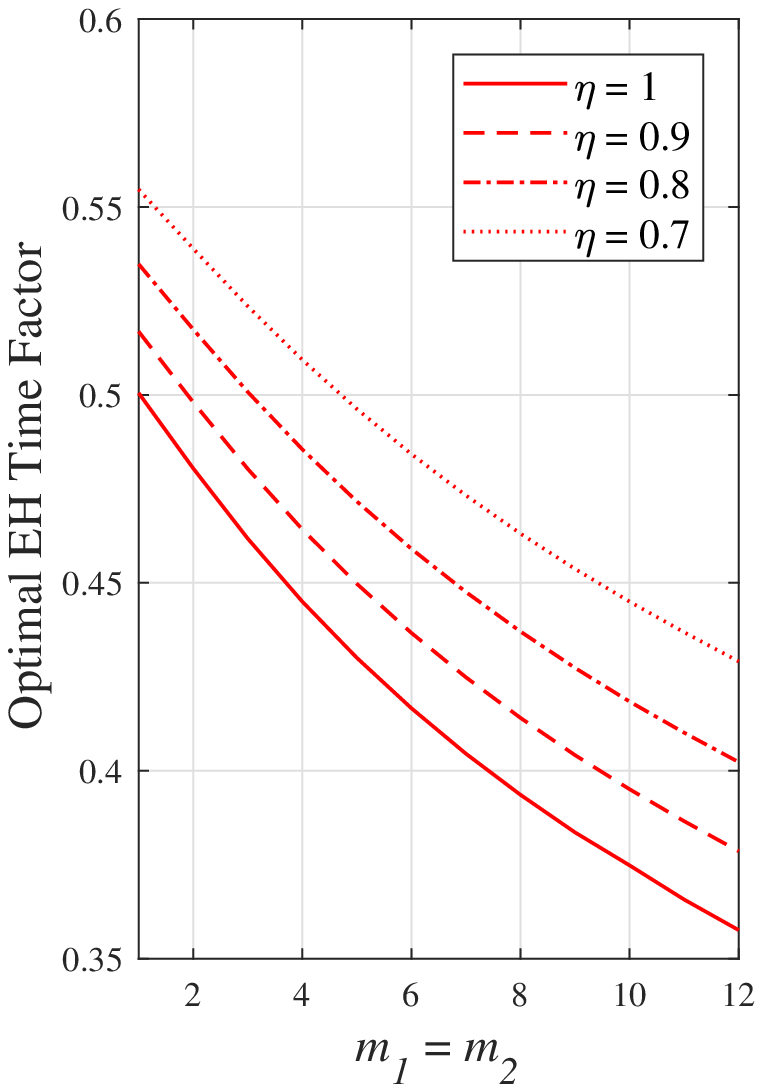}
\par\end{flushleft}

\begin{center}
\caption{\textcolor{black}{\label{fig:Fig4}Optimal EH time factor versus the
fading parameter $m_{i}$ for different values of $\eta$. }}

\par\end{center}%
\end{minipage}
\end{figure}

To illustrate the impact of the loop-back interference channel on
the system performance, we plot in Fig. \ref{fig:Fig3} the ergodic
outage probability as a function of $P_{s}$ for different values
of $\mu_{3}$ when $C_{th}=0.3$bits/s/Hz, $\alpha=0.6$ and $d_{1}=2d_{2}$.
Note that, in this figure, we considered a Nakagami-$m$ fading channel
and kept $\mu_{1}$ and $\mu_{2}$ fixed at 5. It can be seen that
as we improve the loop-back interference channel, i.e., increasing
$\mu_{3}$, the performance deteriorates. It is also noticeable that
the performance enhances as the transmit power is increased and worsens
with increasing the end-to-end distanc\textcolor{black}{e. Furthermore,
Fig. \ref{fig:Fig4} depicts some numerical results of the optimal
EH time factor versus the fading parameter $m_{i}$, $i\in\left\{ 1,2\right\} ,$
with different values of $\eta$ when $P_{s}=1$W and $m_{3}=3$.
Clearly, increasing $m_{i}$ and/or $\eta$ minimizes the outage probability.
}

\section{\label{sec:Conclusion}Conclusion }

This letter analyzed the performa\textcolor{black}{nce of a FD-DF
EH-enabled relaying network over $\kappa$-$\mu$ fading channels.
Accurate mathematical expressions were derived for the ergodic outage
probability. Three special cases of the $\kappa$-$\mu$ fading were
investigated, namely, Rice, Nakagami-$m$ and Rayleigh. Using the
derived exp}ressions, the impact of several system parameters were
examined such as the fading parameters, loop-back interference channel,
end-to-end distance and source transmit power. 
\bibliographystyle{ieeetr}
\bibliography{bib}

\end{document}